\documentclass[preprint,showpacs,preprintnumbers,amsmath,amssymb,superscriptaddress]{revtex4}

\usepackage{graphicx,amsfonts}
\usepackage{epsfig}
\usepackage{dcolumn}
\usepackage{bm}
\begin{document}

\title{The Higgs decay rate to two photons in a model with two fermiophobic-Higgs doublets 
}


\author{H. C\'{a}rdenas}%
\email{hjcardenasr@unal.edu.co}
\affiliation{Departamento de F\'{\i}sica, Universidad Nacional de Colombia, Bogot\'a,
Colombia
}
\author{A. C. B. Machado}%
\email{ana@ift.unesp.br}
\affiliation{
Centro de Ci\^encias Naturais e Humanas,
Universidade Federal do ABC, Santo Andr\'e-SP, 09210-170\\Brazil.
}
\author{V. Pleitez}%
\email{vicente@ift.unesp.br}
\affiliation{
Instituto  de F\'\i sica Te\'orica--Universidade Estadual Paulista \\
R. Dr. Bento Teobaldo Ferraz 271, Barra Funda\\ S\~ao Paulo - SP, 01140-070,
Brazil
}
\author{J.-Alexis Rodriguez}%
\email{jarodriguezl@unal.edu.co}
\affiliation{Departamento de F\'{\i}sica, Universidad Nacional de Colombia, Bogot\'a,
Colombia
}

\date{06/12/12}
%
\begin{abstract}

We consider a  three Higgs doublet model with an $S_3$ symmetry in
which beside the SM-like doublet
there are two fermiophobic doublets. Due to the new charged scalars there
is an enhancement in the two-photon
decay while the other channels have the same decay widths that the SM neutral
Higgs. The fermiophobic scalars are mass
degenerated unless soft terms breaking the $S_3$ symmetry are added.

\end{abstract}


\maketitle

\section{Introduction}
\label{sec:intro}

Recently a new resonance  which is compatible with the Higgs boson of the standard model (SM) with
a mass of 125 GeV has been discovery at the LHC~\cite{cmsatlas}.  As is well known in the context of
that model, nothing constraints the number of fermion generations although,
since the LEP data, we know that there exist only three sequential generations of quarks and leptons.
This triplication may also exist in the scalar sector since here again, nothing constraints the number of
Higgs scalar multiplets and, in particular, the number of Higgs doublets is a free parameter in the model,
although one of them is enough to accommodate vector and fermion masses and their mixing.
In this vain  the multi-Higgs extensions of the standard model are among the most motivated new physics scenarios.
Generally these models have scalar mediated flavor changing neutral currents. Even in the simplest case,
the two-Higgs doublet models, have several possibilities to control those effects~\cite{branco}.
Three-Higgs doublet models~\cite{pakvasa} have not been considered with the same details as those of the two-Higgs doublet case.
This is not a surprise since in this case the analysis of the scalar potential is much more complicated.
However, discrete symmetries may simplified the scalar potential, for example the $A_4$ symmetry has been considered
in Ref.~\cite{tria4}. Recently, it was shown that the $S_3$ symmetry is very efficient to constraint the
scalar potential allowing to obtain the mass spectra and the matrix which diagonalize the mass square matrices~\cite{flavor}.
The symmetries for the two-Higgs doublet model was obtained in Refs.~\cite{symmetries2hd} and in the case of three-Higgs doublet
models in \cite{symmetries3hd}.

If one or more extra Higgs doublets do exist in Nature it seems that their existence is due to some reason that allows
to explain something else that could not be explained by the minimal model, i.e., with only one Higgs doublet.
On one hand, it is possible that extra scalars may explain the mass spectra and mixing in the fermion sectors~\cite{bhatta}, or
on the other hand, it is possible that the extra Higgs doublets may help to understand the observed dark matter. In the latter
case the extra Higgs bosons have to be of the fermiophobic type~\cite{inert}. The fermiophobic Higgs boson is defined as: all the fermion couplings to the Higgs boson are set to zero and the bosonic couplings are the same as in the standard model. This in fact has been already considered in the case of one fermiophobic doublet-Higgs  model~\cite{dark}. Here we will
consider the LHC phenomenology of the three doublet model with $S_3$ symmetry~\cite{ishimori} which was put forward in
Ref.~\cite{flavor} and which has two fermiophobic doublets.

The outline of this paper is as follows: In the next section  we review the main 
feature of the three scalars model. In Sec. III we give the interactions of the model. In 
Subsec.~IIIA the Yukawa interactions, in Subsec.~IIIB the gauge interactions while in 
Subsecs.~IIIC and D we write down explicitly the trilinear interactions.  In Sec. IV we 
show the decay rate into two photon of the SM-like neutral scalar. We devote Sec.~V 
for our conclusions.  In the Appendix A we show how the mass spectra are modified if 
we add soft terms to the scalar potential.
\section{The scalar sector}
\label{sec:scalar}

Let us consider an extension of the SM electroweak theory which consists in adding two extra scalars, $SU(2)_L$
doublets, with $Y=+1$.  The three scalar doublets are in a singlet $S$ and a doublet $D$ of $S_3$.
The $SU(2)_L \otimes U(1)_Y \otimes S_3$ invariant scalar potential is given by
\begin{equation}
V=V(D,S)+V(D,S)_{soft},
\label{potential1}
\end{equation}
where:
\begin{eqnarray}
V(D,S) &=& \mu^2_sS^\dagger S+\mu^2_d [D^\dagger\otimes  D]_1  +\lambda_1
([D^\dagger\otimes  D]_1)^2
+  \lambda_2 [(D^\dagger\otimes D)_{1^\prime}(D^\dagger\otimes
D)_{1^\prime}]_1
\nonumber \\ &+&\lambda_3[(D^\dagger \otimes D)_2(D^\dagger\otimes D)_2]
+\lambda_4(S^\dagger S)^2+
\lambda_5[D^\dagger\otimes D]_1 S^\dagger  S+\lambda_6 S ^\dagger[D^\dagger
\otimes D]_1 S
\nonumber \\ &+&
\{ \lambda_7[(S^\dagger \otimes D)_2(D^\dagger \otimes
S)_2]_1+\lambda_8[(S^\dagger\otimes D)_2(D^\dagger \otimes D)_2]_1+H.c.\}
\label{potential1}
\end{eqnarray}
and $V_{soft}$ denote soft terms breaking $S_3$ symmetry explicitly, see Ref.~\cite{flavor}. The effects of the soft terms on the scalar masses
are considered in the Appendix~\ref{sec:softterms}.

There are two ways to build the singlet $S$ and the doublet $D$ which are not equivalent. In the first one, we call model A,
the reducible triplet representation of the discrete symmetry~$S_3$~:~$\textbf{3}=(H_1,H_2,H_3)$ with the usual notation
$H_i=(H^+\,H^0_i)^T$ in which $H^0_i = (1/\sqrt2)(v+\eta^0 + i A^0)$. This reducible representation is the direct sum of one singlet and one doublet $S_3~=
~\textbf{2}+\textbf{1}\equiv D+S$, where $S$ and $D$ are give by
\begin{eqnarray}
&&S=\frac{1}{\sqrt3}(H_1+H_2+H_3)\sim\textbf{1},\nonumber \\&&
D\equiv (D_1,D_2)=\left[\frac{1}{\sqrt6}(2H_1-H_2-H_3),\frac{1}{\sqrt2}(H_2-H_3)\right]\sim\textbf{2},
\label{ma}
\end{eqnarray}
and the other way, we denote model B, is such that
\begin{equation}
S = H_1\sim\textbf{1} , \quad D = (H_2, H_3) \sim\textbf{2}.
\label{ma2}
\end{equation}
We also impose a vacuum alignment in each case: $(v,v,v)$ and $(v_{SM},0,0)$ in model A and B, respectively. This vacuum
alignment give a global and stable minimum of the scalar potential if other conditions are satisfied too~\cite{flavor}. In
both cases the constraint equations reduce to $\mu^2_s=-\lambda_4v^2_{SM}$ which implies that $\lambda_4>0$. The difference
is that $3v^2=v^2_{SM}$ in the model A, and $v^2_1=v^2_{SM}$, in model B, see Ref.~\cite{flavor} for details.

With the scalar potential in Eq.~(\ref{potential1}) in model A the mixing matrix in all the scalar, pseudoscalar and
charged scalar sectors is given by the tribimaximal matrix
\begin{equation}
U_{TBM} =\left(
\begin{array}{ccc}
\frac{1}{\sqrt3} & - \sqrt{\frac{2}{3}} & 0\\
\frac{1}{\sqrt{3}}& \frac{1}{\sqrt{6}}& - \frac{1}{\sqrt{2}}\\
\frac{1}{\sqrt{3}}& \frac{1}{\sqrt{6}} & \frac{1}{\sqrt{2}}
\end{array}
\right),
\label{tribi}
\end{equation}
and the masses are the following: in the $CP$ even sector
\begin{eqnarray}
m^2_{h_1}=\lambda_4v^2_{SM}, \qquad
m^2_{h_2}=m^2_{h_3}=\mu^2_d+\frac{1}{2}\bar{\lambda}^\prime v^2_{SM},
\label{mrs1}
\end{eqnarray}
where  $\bar{\lambda}^\prime=\lambda_5+ \lambda_6+2\lambda_7$,  and denoting as $h^0_i$ the mass eigenstates, we
have $\eta^0_i=(U_{TBM})_{ij}h^0_j$, where $U_{TBM}$ is given in (\ref{tribi}). The scalar $h^0_1$ can be identified
with the standard model Higgs scalar.

In the CP-odd neutral scalars sector, we obtain the following masses:
\begin{eqnarray}
   m^2_{a_1}=0, \qquad
m^2_{a_2}=m^2_{a_3}=\mu^2_d+\frac{1}{2} \bar{\lambda}^\prime v^2_{SM}
\label{mps1}
\end{eqnarray}
Denoting $a^0_i$ the pseudo-scalar mass eigenstates, we have $A^0_i=(U_{TBM})_{ij}a^0_j$.

Similarly in the charged scalars sector we obtain
the following masses:
\begin{eqnarray}
m^2_{c_1}=0,\qquad
m^2_{c_2}=m^2_{c_3}=\frac{1}{4}(2\mu^2_d+\lambda_5v^2_{SM}),
\label{mcs1}
\end{eqnarray}
and denoting $H^+_i$ denote the charged scalar symmetry eigenstates and $h^+_i$ the respective mass eigenstates,
we have $H^+_i=(U_{TBM})_{ij}h^+_j$.

In model A the $SU(2)$ doublets can be written in terms of the mass eigenstates using the mixing matrix em Eq.(\ref{tribi}), resulting in:
\begin{eqnarray}
&&S=\left(\begin{array}{c}
h^+_1
\\
\frac{1}{\sqrt2}(3v+h^0_1+ia^0_1)
\end{array}
\right),\nonumber \\&&
D_1 = -\left(\begin{array}{c}
h^+_2 \\
\frac{1}{\sqrt2}(h^0_2+ia^0_2 )\end{array}\right),\;\;
D_2 =  -\left(\begin{array}{c}
h^+_3 \\
\frac{1}{\sqrt2}(h^0_3+ia^0_3 )\end{array}\right).
\label{casoa}
\end{eqnarray}

However, in model B the mass matrices are diagonal, i.e. there is no mixing in each charge
sector. In general, the eigenvalues are equal to those in Eq.(\ref{mrs1}) for $CP$ even sector, Eq.(\ref{mps1}) for $CP$ odd sector  and Eq.(\ref{mcs1}) for the charged scalar sector, respectively.

Notice that the mass degeneracy in the fermiophobic sector is a prediction of the $S_3$ symmetry but there may be accidental mass degeneracy with the SM-like
Higgs boson too.

The possibility that two mass degenerated Higgs bosons with mass near the 125 GeV has been discussed in 
literature~\cite{ellwanger,gunion1,gunion2,gunion3,haber}. The main difference with the present model is that two of the
Higgs doublets are fermiophobic, they do not interact with quarks or leptons at tree level. On the other hand, they can be produced in
accelerators like the LEP by the Higgstrahlung mechanism $e^+e^-\to Z^*\to ZX$ or in hadronic colliders $qq^\prime\to VV\to X$ where $X$ denotes any 
neutral scalar. Moreover, since  they are fermiophobic scalars they do not decay into fermions and they behave as invisible Higgses. Bounds on the masses of 
femiophobic Higgs boson in the diphoton decay channel exclude this sort of scalars in the ranges 110-118 GeV and 119.5 and 121 GeV~\cite{atlas}.
This is the case of the fermiophobic Higgs in the present model. Moreover, the decay $ZZ\to 4l$ is exactly the same as in the SM since only one
of the neutral scalar (the one which is not fermiophobic, $h_1$) contributes to these decays.

\section{Interactions}

\subsection{The Yukawa sector}
\label{subsec:yukawa}
The Yukawa interactions are equal in both models when vacua are aligned as before. Only one of the doublets interacts with quarks and leptons and
the other two are fermiophobic doublets.
In the lepton sector all lepton fields transform as singlet under $S_3$ and for these reason they only interact with the singlet $S$:
\begin{equation}
-\mathcal{L}_l=\bar{L^\prime}_{iL}G^l_{ij} Sl^\prime_{jR}+\bar{L^\prime}_{iL}G^\nu_{ij}\tilde{S}\nu^{\,\prime}_{jR} +H.c..
\label{lmasses}
\end{equation}
where the prime fields denote symmetry eigenstates which are written in terms of the mass (unprimed) fields by using unitary matrices:
\begin{equation}
l^\prime_{iL}= (U^l_L)_{ij} l_{jL} \;\; , \;\; l^\prime_{iR}= (U^l_R)_{ij} l_{jR} \;\; , \;\; \nu_{iL}^{\,\prime}= (U^\nu_L)_{ij} \nu_{jL}
\;\; , \;\; l^\prime_{iR}= (U^l_R)_{ij} l_{jR}
\label{rot1}
\end{equation}

The Yukawa interactions written in terms of the mass eigenstates are:
\begin{eqnarray}
-\mathcal{L}_l & = &\bar{\nu}_{iL}\frac{\hat{M}^l_{i}}{v_{SM}} (V_{PMNS})_{ij}
 l_{jR}h^+_1 +  \bar{l}_{iL} \frac{\hat{M}^l_{i}}{v_{SM}}l_{jR}
 \left[1 +  \frac{h^0_1 + ia^0_1 }{\sqrt2} \right]
\nonumber \\&+&
\bar{l}_{iL}\frac{\hat{M}^\nu_{i}}{v_{SM}} (V_{PMNS})_{ij}
 \nu_{jR}h^-_1  +  \bar{\nu}_{iL} \frac{\hat{M}^\nu_{i}}{v_{SM}}\nu_{iR}
 \left[1+  \frac{h^0_1 +ia^0_1 + }{\sqrt2}\right]
 +H.c.,
\label{lnt3}
\end{eqnarray}
where we have defined $V_{PMNS}=U^{l\dagger}_LU^\nu_L$.

Similarly, all quarks fields are singlet under $S_3$, hence as in the lepton case, they only interact with the singlet $S$:
\begin{equation}
-\mathcal{L}_q=\bar{Q^\prime}_{iL}G^u_{ij}\tilde{S}u^\prime_{jR}+\bar{Q^\prime}_{iL}G^d_{ij}Sd^\prime_{jR} +H.c.,
\label{qmasses}
\end{equation}
and using
\begin{equation}
u^\prime_{iL}= (U^u_L)_{ij} u_{jL} \;\; , \;\; u^\prime_{iR}= (U^u_R)_{ij} l_{jR} \;\; , \;\; d_{iL}^{\,\prime}= (U^d_L)_{ij} d_{jL}
\;\; , \;\; d^\prime_{iR}= (U^d_R)_{ij} d_{jR}
\label{rot2}
\end{equation}
we write the Yukawa interactions in terms of the quark mass eigenstates
\begin{eqnarray}
-\mathcal{L}_q & = &\bar{u}_{iL}\frac{\hat{M}^d_{i}}{v_{SM}} (V_{CKM})_{ij}
  d_{jR} h^+_1+  \bar{d}_{iL} \frac{\hat{M}^d_{i}}{v_{SM}}\left[1
 + \frac{h^0_1  }{\sqrt2}  \right] d_{iR}
\\ \nonumber &+&
\bar{d}_{iL}\frac{\hat{M}^u_{i}}{v_{SM}} (V_{CKM})_{ij}
  u_{jR}h^-_1 +  \bar{u}_{iL} \frac{\hat{M}^u_{i}}{v_{SM}}u_{iR}\left[1+
  \frac{h^0_1 }{\sqrt2} \right]+H.c.,
\label{lnt32}
\end{eqnarray}
where we have defined $V_{CKM}=U^{u\dagger}_LU^d_L$. Above $\hat{M}$ denotes diagonal mass matrices in the respective charge sector.

As in the standard model the masses and the $V_{CKM}$ and $V_{PMNS}$ mixing matrices can be accommodated but their values are not explained.

\subsection{Gauge-scalar interactions}
\label{subsec:a2}

In this sector, when the scalar doublets are written in terms of the mass eigenstates, only one of the scalar doublets contribute to the vector boson masses as in the SM.
The $SU(2)_L\otimes U(1)_Y\otimes S_3$ invariant gauge interactions are
\begin{eqnarray}
\mathcal{L}_{gauge}&=&(\mathcal{D}_\mu S)^\dagger (\mathcal{D}^\mu S)+(\mathcal{D}_\mu D)^\dagger (\mathcal{D}^\mu D)
\nonumber \\&=&(\mathcal{D}_\mu H_1)^\dagger (\mathcal{D}^\mu H_1)+(\mathcal{D}_\mu H_2)^\dagger (\mathcal{D}^\mu H_2)+
(\mathcal{D}_\mu H_3)^\dagger (\mathcal{D}^\mu H_3),
\label{gaugema}
\end{eqnarray}
where $S, D$ or $H_i$ are symmetry eigenstates. Using the first line and the fields in Eq.~(\ref{casoa}) and (\ref{casob}), we can write the Higgs scalar gauge interactions
in terms of the mass eigenstates:
\begin{eqnarray}
\mathcal{L}_{gauge}&=&(\mathcal{D}_\mu h_1)^\dagger (\mathcal{D}^\mu h_1)+(\mathcal{D}_\mu h_2)^\dagger (\mathcal{D}^\mu h_2)
+(\mathcal{D}_\mu h_3)^\dagger (\mathcal{D}^\mu h_3).
\label{a141}
\end{eqnarray}
where $h_i=[h^+_i,\;(h^0_i+ia^0_i)/\sqrt2]^T,\;i=1,2,3$ are the $SU(2)$ doublets written in terms of the mass eigenstates. We have omitted the mass term, i.e.,
the VEV in $h_1$. The covariant derivative $\mathcal{D}_\mu$ is the same of the standard model.

\subsection{Trilinear Interactions in model A}
\label{subsec:trilineara}

The trilinear interactions in model A with or without the soft terms (see Appendix~\ref{sec:softterms}) are as follows
\begin{equation}
\frac{v_{SM}}{2\sqrt3}\,\left[\lambda_4h^-_1h^+_1+\lambda_5(h^-_2h^+_2+h^-_3h^+_3)\right](h^0_1-ia^0_1).
\label{a11}
\end{equation}

In the same way for the second scalar,
we have
\begin{equation}
\frac{v_{SM}}{2\sqrt3}\,\left[(\lambda_6+\lambda_7)h^-_1h^+_2-\lambda_8(h^-_2h^+_2+h^-_3h^+_3)\right](-h^0_2+ia^0_2),
\label{a12}
\end{equation}
Note that the vertex with $h^-_1h^+_3$ does not exist. Finally, for the third scalar
\begin{equation}
\frac{v_{SM}}{2\sqrt3}\,\left[-(\lambda_6+\lambda_7)h^-_1h^+_3+\lambda_8h^-_2h^+_3\right](-h^0_3+ia^0_3),
\label{a13}
\end{equation}
and in this case, the vertex with $h^-_1h^+_2$ that does not exist.

In the neutral scalar and pseudo-scalar sector we have (up to a factor
$v_{SM}/2\sqrt3$)
\begin{eqnarray}
&&[\lambda_4(h^0_1h^0_1+a^0_1a^0_1)+\lambda_5(h^0_2h^0_2+h^0_3h^0_3+a^0_2a^0_2+
a^0_3a^0_3)](h^0_1-ia^0_1)\nonumber \\&&
-[(\lambda_6+\lambda_7)(h^0_1+ia^0_1) ( h^0_2+ia^0_2)
+\lambda_8(h^0_2h^0_2+ a^0_2a^0_2 -
h^0_3h^0_3-a^0_3a^0_3)](-h^0_2+ia^0_2)\nonumber \\&&
-[(\lambda_6+\lambda_7)(-h^0_1h^0_3+a^0_1a^0_3+
i(a^0_1h^0_3-a^0_3h^0_1))-
2\lambda_8(a^0_2a^0_3+h^0_2h^0_3)](-h^0_3+ia^0_3),
\label{a14}
\end{eqnarray}

\subsection{Trilinear Interactions in model B}
\label{subsec:trilinearb}

\subsubsection{Without the soft terms}
\label{subsubsec:b1}

In model B without the soft terms we have the following trilinear
interactions:
\begin{equation}
\frac{v_{SM}}{2}\,\left[\lambda_4h^-_1h^+_1+\lambda_5(h^+_2h^-_2+h^+_3h^-_3)\right](h^0_1-ia^0_1),
\label{a21}
\end{equation}
\begin{equation}
\frac{v_{SM}}{2\sqrt2}\,\left(\lambda_6+\lambda_7\right)h^-_1h^+_2(-h^0_2+ia^0_2),
\label{a22}
\end{equation}
and
\begin{equation}
\frac{v_{SM}}{2\sqrt2}\,\left(\lambda_6+\lambda_7\right)h^-_1h^+_3(h^0_3+ia^0_3).
\label{a23}
\end{equation}

In the neutral sector (up to a factor $v_{SM}/2$)
\begin{eqnarray}
&&\left[\frac{\lambda_4}{2}(h^0_1h^0_1+a^0_1a^0_1)+(\lambda_5+\lambda_6+\lambda_7)(h^0_2h^0_2+
h^0_3h^0_3+a^0_2a^0_2 +a^0_3a^0_3)\right]
(h^0_1-a^0_1)\nonumber \\&&
+\frac{1}{4}(\lambda_6+\lambda_7)(h^0_1+ia^0_1)(h^0_2-ia^0_2)
(h^0_2+ia^0_2)\nonumber \\&&
+\frac{1}{4}(\lambda_6+\lambda_7)(h^0_1+ia^0_1)(h^0_3-ia^0_3)(h^0_3+ia^0_3)
\label{a241}
\end{eqnarray}

\subsubsection{With soft terms}
\label{subsubsec:b2}

In model B (see Appendix \ref{sec:softterms})   when the soft terms are included we have the following trilinear
interactions:
\begin{equation}
\frac{v_{SM}}{2}\,\left[\lambda_4h^-_1h^+_1+\lambda_5(h^+_2h^-_2+h^+_3h^-_3)\right](h^0_1+ia^0_1),
\label{a21st}
\end{equation}
\begin{equation}
\frac{v_{SM}}{2\sqrt2}\,\left(\lambda_6+\lambda_7\right)h^-_1(h^+_3-h^+_2)(-h^0_2+ia^0_2),
\label{a22st}
\end{equation}
\begin{equation}
\frac{v_{SM}}{2\sqrt2}\,\left(\lambda_6+\lambda_7\right)h^-_1(h^+_3+h^+_2)(-h^0_3-ia^0_3),
\label{a23st}
\end{equation}
and, up to a factor $v_{SM}/\sqrt2$
\begin{equation}
\frac{v_{SM}}{2\sqrt2}\left[\lambda_4(h^0_1h^0_1+a^0_1a^0_1)+2(\lambda_5+\lambda_6+\lambda_7)(h^0_2h^0_2+
h^0_3h^0_3+a^0_2a^0_2 +a^0_3a^0_3)\right]
(h^0_1-a^0_1)
\label{a24st}
\end{equation}

We see that model A differs from the model B only in the trilinear (and
quartic but we have not shown they here) interactions. Model B
also has difference scalar-scalar interactions depending if we add or not,
the soft term to the scalar potential. Thus, those possibilities
may be distinguished when Higgs self-couplings are measured at the
LHC~\cite{self}.



\begin{figure}[ht]
\begin{center}
\includegraphics[width=3.5in]{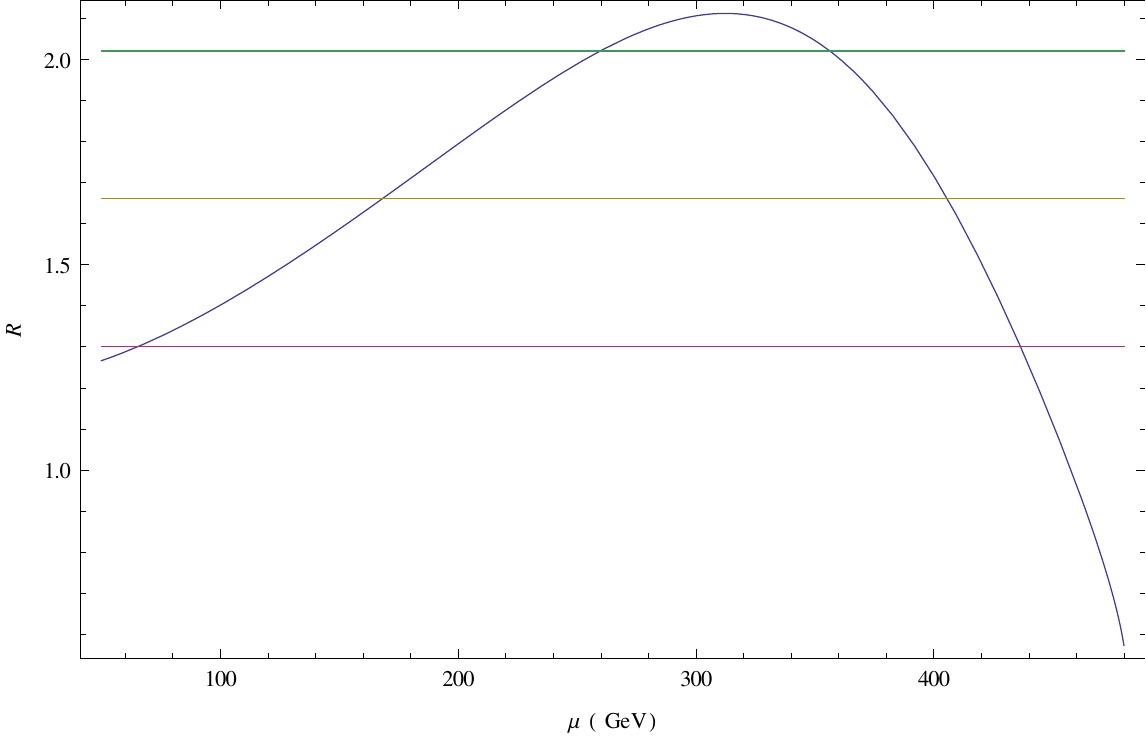}
\caption{The reduced branchig fraction $R_{\gamma \gamma}$ vs $\mu_d$ using 
$\lambda_5=1$ and $\lambda_6=\lambda_7=0$ in the model A. The solid lines 
correspond to the combined experimental value $R_{\gamma \gamma}=1.66 \pm 0.36$ \cite{Belanger}. An excluded area is found around 260-355 GeV.}
\label{gf1}
\end{center}
\end{figure}

\section{Results and discussions}

\begin{figure}[ht]
\begin{center}
\includegraphics[width=3.5in]{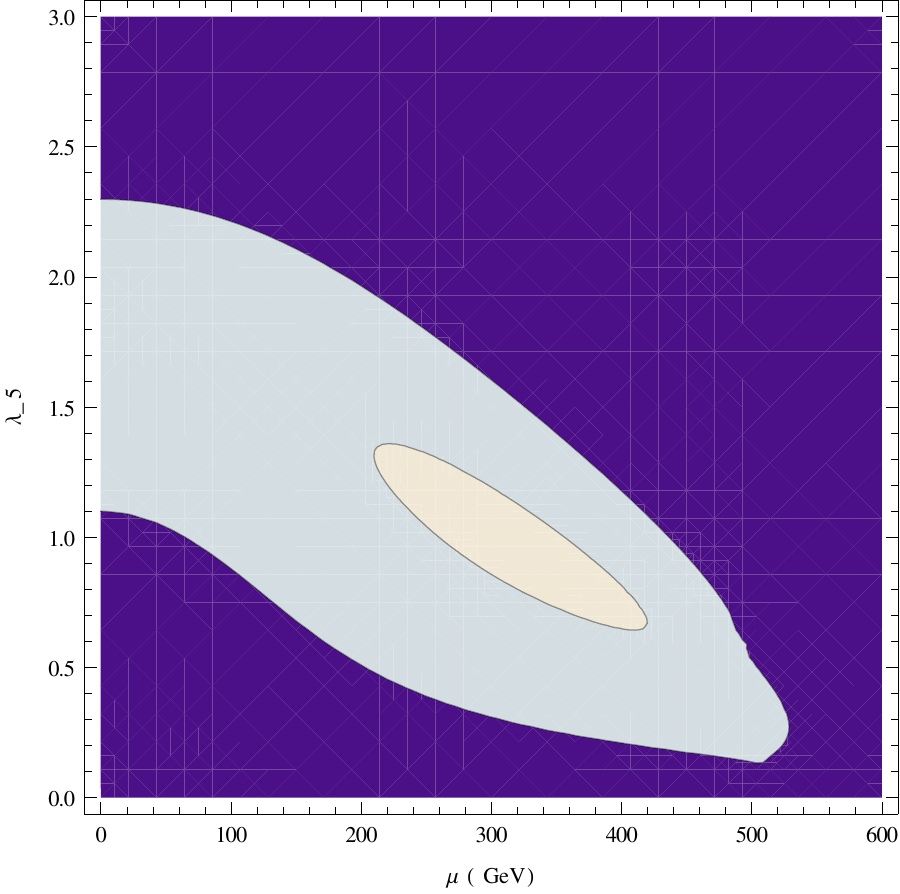}
\caption{Contour plot $\lambda_5$-$\mu_d$ with $\lambda_6=\lambda_7=0$ constrained by $R_{\gamma \gamma}=1.66 \pm 0.36$. The allowed region is the ligth region with
a white excluded area inside. Some points of this plot are shown explicitly in table 1 with their associated Higgs boson masses.}
\label{g2}
\end{center}
\end{figure}

We can explore the phenomenology associated to the Higgs sector of this model under some basic assumptions. We are going to consider only the model A in this section without
soft terms added although they do not modify our results.
If $h^0_1$ is the SM-like Higgs boson, it mass has to be near $125$ GeV, it implies
$\lambda_4=0.26$.
From Eqs.~(\ref{mrs1})-(\ref{mcs1}) we obtain typical values of the scalar bosons masses in both models, respectively. We can evaluate the decay channels of the neutral CP-even Higgs $h^0_1$ in a mass range around $125$ GeV and we can compare the branching fractions with the SM results. The Higgs sector depends on the
Higgs mass spectrum which is parameterized in terms of $\lambda_5$, $\lambda_6$, $\lambda_7$, $\mu_d^2$ in the model A. On the other hand, the fermiophobic Higgs fields $h_{2,3}$
only interact throught the trilinear terms already mentioned and if we assume these Higgs bosons with a mass bigger than $125$ GeV then they only are going to contribute
to the $h\to \gamma \gamma$ and $h \to \gamma Z$. It is interesting the $h\to \gamma \gamma$ decay because there is an excess of events above the SM predictions. We are going
to focus on this decay mode because in the other decay channels there are not significant contributions respect to the SM expectations.
It is
useful to define a reduced signal rate $R$ relative to the expected signal of the SM Higgs boson~\cite{Belanger}
\begin{equation}
 R_{\gamma \gamma}= \frac{\sigma(pp \to h_1^0)}{\sigma(pp \to h_{SM})} \frac{BR(h_1^0 \to \gamma \gamma)}{BR(h_{SM} \to \gamma \gamma)},
\end{equation}
\begin{table}
 \centering
\begin{tabular}{|c|c|c|c|}
\hline\hline
$\lambda_5$ & $\mu_d$ (GeV) & $m_{h_2}$ (GeV)&$m_{c_2}$ (GeV)  \\ \hline
1 &0&175&123 \\ \hline
2.3 & 0 & 264 & 186  \\ \hline
1.3 & 203 & 283 & 200  \\ \hline
0.6 & 410 & 433& 306  \\ \hline
0.11& 511 & 514 &363  \\ \hline
\end{tabular}
\caption{Some points from the figure 2 and the associated Higgs boson masses. As we can see from Eqs.(\ref{mrs1}) and (\ref{mps1}) we have $m_{h_2} = m_{a_2}$.}
\label{tabla2}
\end{table}
where the first factor is associated with the production mechanism which in our case is mainly throught the gluon-gluon fusion and the second factor is the reduced
branching fraction for the channel under consideration. In model A, the first factor will be one because there are not any new contribution from the fermiophobic Higgs bosons
interactions to the Higgs production, the new Higgs bosons do not couple to the quarks. Therefore, $R_{\gamma \gamma}$ is the reduced branching fraction. 
In the $h_1^0 \to \gamma \gamma$ decay channel there are contributions in the loop from the couplings $h_1^0 h_{2,3}^{+} h_{2,3}^{-}$ which are proportional to $\lambda_5$. 
There are experimental reports from CMS and ATLAS collaborations to the $R_{\gamma \gamma}$ fraction in the $\gamma \gamma$ mode and the combined results imply in $R_{\gamma \gamma} =1.66 \pm 0.36$ \cite{Belanger}, which we 
are going to use to constraint the model A parameters. In figure 1, we have plotted the $R_{\gamma \gamma}$ fraction versus the parameter $\mu_d$ using 
$\lambda_5=1$ and $\lambda_6=\lambda_7=0$. The parameters $\lambda_{6,7}$ are involved in the Higgs boson masses while $\lambda_5$ is also appearing in the trilinear 
couplings. From the figure \ref{gf1}, there is an allowed region for $\mu_d$ between $65-260$ GeV and $355-435$ GeV and excluded $260-355$ GeV, these intervals correspond
to fermiophobic Higgs boson masses of $m_{h_2} = m_{a_2} = 185-312$ GeV and $m_{c_2}=130-221$ GeV in the first allowed interval and  $m_{h_2} = m_{a_2} = 279-395$ GeV and
$m_{c_2}=279-331$ GeV  in the second one. Here we should emphasize that in the model A the fermiophobic Higgs fields are mass degenerate. In figure \ref{g2}, 
we have make a contour plot in the plane $\lambda_5$-$\mu_d$ using the experimental value of the reduced branching fraction $R_{\gamma \gamma}$ in order to
explore the space parameter of  $\lambda_5$. The allowed region is the light colored region and there are excluded areas around 
and in the middle of the contour which is the white area. Some Higgs boson masses gotten from the figure \ref{g2} are in table \ref{tabla2}. A brief comment about the 
parameters $\lambda_{6,7}$ is that their values are not affecting the regions obtained because they only appear in the expression of the fermiophobic Higgs boson masses in the loop.

A brief comment about the production of the fermiophobic Higgs bosons should be addressed. The recent discovery of a Higgs like boson at the LHC does not rule out the possibility
of a Higgs boson decaying into a channel with invisible decay products as in our case the $h_1^0$ into $h_{2,3}^0 h_{2,3}^0$ or $h_{2,3}^\pm h_{2,3}^\pm$ \cite{invisible}.  The most important
channel for the detection of this invisible modes is vector boson fusion since it has a large cross section but also it has a large systematic uncertantities and it is difficult
to estimate the QCD background. Another option is the associated production channel $W h^0$ or $Zh^0$, however the $Wh^0$ channel is diluted by the inclusive $W$ backgroud which
makes it difficult to analyze instead the $Zh$ channel is more promising \cite{invisible}. On the other hand, this fermiophobic Higgs bosons are already candidates to dark matter and constraints 
from their production through channels like $h_{2,3}^0 h_{2,3}^0 \to h_1^0 \to \gamma  \gamma$ will be expected.

\section{Conclusions}

Fermiophobic Higgs fields only interacts with itself and to
to other scalar and vector bosons or active multiplets, they have been called
inert~\cite{barbieri} or dark~\cite{cao}
Higgses and they may transform under the gauge symmetries of the SM
in non-trivial way as doublets~\cite{inert,barbieri}, or in a trivial way,
i.e.,
singlets~\cite{inert2}. They have been considered as solutions to the
hierarchy problem or/and as a good cold dark matter
candidates~\cite{barbieri,inert2}. 
Here, we have build up an extension of the SM adding two extra doublet scalar and using a $S_3$ symmetry. There are
two ways to build the singlet and the doublet of $S_3$, we are called them model A and B (see Appendix \ref{sec:softterms}).
The models are different in their trilinear and cuartic interactions but they have a dark degenerate scalar spectrum. The interactions
of the SM-like Higgs boson are identical to the SM. Therefore, the only effect from the dark scalars is in the one loop processes like
the Higgs boson into two photon decay. We have evaluated the reduced fraction $R_{\gamma \gamma}$ in order to get constraints for the
parameters $\lambda_5$ and $\mu_d$ of the model A. On the other hand, models A and B are predicting the same $h_1 \to \gamma \gamma$ width decay
but they are not equivalent in the invisible modes.
In general a fermiophobic neutral scalar
decays to $W $and $Z$ bosons proceeds as in the SM, while the decay to
photons proceeds via $W,h^{\pm}$ loops, since its
decays to photons via fermion loops are excluded. If this were the case
for the doublet of the standard model i.e.,
if the fermion masses have a different origin from that of the gauge boson,
it  is excluded at 95\% confidence level in the mass range $110-194$ GeV,
and at 99\% confidence level in the
mass ranges $110-124.5$ GeV, $127-147.5$ GeV, and $155-180$
GeV~\cite{cms}. Notice that in this case there is a small windows around
124.5-127 GeV.
Hence, the constraints
above are not directly applicable to the model considered in this work that has the usual Higgs doublet plus two dark doublets. However in the present model, fermions masses arise from the Higgs doublet
which also contributes to the gauge boson masses but the latter particles
also have contributions from the dark doublets. 

Note that for the calculations performed in this work we consider that $ \lambda_6 = \lambda_7 = 0$, since they do not contribute in the photon-photon loop, however, as a result of this assumption, the charged boson is lighter  than the neutral boson, some values are shown in Table 1. As a consequence the neutral scalar is not a good candidate for dark matter. However, if we consider that good candidates for dark matter must satisfy the relation $m_c^2 - m_h^2 \geqslant 0$ implying $-\frac{\mu^2}{v_{SM}^2} - \frac12 \lambda_5-(\lambda_6+2 \lambda_7) \geq 0 $, is easy to see that, for example, with $\mu_d = 82$ GeV, $\lambda_5 = 1$, $\lambda_6 = - 0.82$  and $\lambda_7 = 0$ we obtained  $m_h = 110$ GeV and $m_c = 136$, thus we can have a dark matter scenario in these models by imposing that the inequality must be satisfied.

---

\acknowledgments
HC and JAR acknowledge the hospitality of IFT, Sao Paulo were this work was finished.
One of the author (ACBM) was supported by CAPES and (VP) was partially
supported by CNPq and FAPESP. HC and JAR were partially supported by grant 14844 DIB-UNAL.


\appendix

\section{Scalar masses with soft terms in the scalar potential}
\label{sec:softterms}

The mass degeneracy above is due to a residual symmetry that can be broken,
if necessary, by including the soft terms:
\begin{eqnarray}
V(D,S)_{soft} &=& \mu^2_{22}H^\dagger_2H_2 + \mu^2_{33}H^\dagger_3H_3 + \left( \mu^2_{23}H^\dagger_2H_3 +H.c. \right)
\label{potentialsoft}
\end{eqnarray}

In model A, taken into account the soft terms with the condition $\mu^2_{22}=\mu^2_{33}=-\mu^2_{23}\equiv \mu^2>0$, the mass spectrum in Eqs.~(\ref{mrs1})-(\ref{mcs1})
is as follows: only third scalar in each sector becomes heavier since its mass gain a contribution of $\mu^2$
\begin{eqnarray}
&&m^2_{h_1}=m^2_h=\frac{2}{3}\lambda_4v^2_{SM}, \quad
m^2_{h_2}=\mu^2_d+\frac{1}{2}\bar{\lambda}^\prime v^2_{SM}, \quad m^2_{h_3}=\mu^2_d+\frac{1}{2}\bar{\lambda}^\prime v^2_{SM}+\mu^2,\nonumber \\&&
 m^2_{a_1}=0, \quad m^2_{a_2}=\mu^2_d+\frac{1}{6} \bar{\lambda}^\prime v^2_{SM}\quad m^2_{a_3}=\mu^2_d+\frac{1}{6} \bar{\lambda}^\prime v^2_{SM}+\mu^2
 \nonumber \\&&
 m^2_{c_1}=0,\quad
m^2_{c_2}=\frac{1}{2}\mu^2_d+\frac{\lambda_5}{12}v^2_{SM},\quad m^2_{c_3}=\frac{1}{2}\mu^2_d+\frac{\lambda_5}{12}v^2_{SM}+\mu^2,
\label{numma2}
\end{eqnarray}
and the mass degeneracy in the iner sector has been broken but it is still possible and accidental degeneracy with the SM-like Higgs scalar.

The mixing matrix remains the same as in Eq.~(\ref{tribi}).

In model B, when the soft terms are included with the condition
$ \mu^2_{22} =\mu^2_{33} = \nu^2$, and $\mu^2_{23}  = \mu^2$, we have
\begin{eqnarray}
&&\bar{m}^2_{h_1}=\lambda_4v^2_{SM},\;\; \bar{m}^2_{h_2}=\mu^2_d+\frac{1}{2}\bar{\lambda}^\prime v^2_{SM}+2\mu^2-\nu^2,\;\;
\bar{m}^2_{h_3}=\mu^2_d+\frac{1}{2}\bar{\lambda}^\prime v^2_{SM}+2\mu^2+\nu^2,\nonumber \\&&
\bar{m}^2_{a_1}=0,\;\; m^2_{a2}=\mu^2_d+\frac{1}{2}\bar{\lambda}^\prime v^2_{SM}+2\mu^2-\nu^2, \;\;m^2_{a3}=
\mu^2_d+\frac{1}{2}\bar{\lambda}^\prime v^2_{SM}+2\mu^2+\nu^2,\nonumber \\&&
\bar{m}^2_{c_1}=0,\;\;m^2_{c_2}=\frac{1}{4}(2\mu^2_d+\lambda_5v^2_{SM})\!\!+\mu^2\!\!-\frac{1}{2}\nu^2,\;\; m^2_{c_3}=\frac{1}{4}(2\mu^2_d+
\lambda_5v^2_{SM})\!+\!\mu^2\!+\!\frac{1}{2}\nu^2,
\label{massfb2}
\end{eqnarray}
and the mixing matrix between the respective components of $H_2$ and $H_3$ is
\begin{equation}
U = \left(
\begin{array}{ccc}
1 &                                               0 & 0 \\
0 &-\frac{1}{\sqrt{2}}  & \frac{1}{\sqrt{2}} \\
0 &  \frac{1}{\sqrt{2}} &  \frac{1}{\sqrt{2}}
\end{array}
\right),
\label{uva}
\end{equation}
and the mixing between $H_2$ and $H_3$ sector is maximal. In this case $S$ is still as in Eq.~(\ref{casoa}) but now
\begin{eqnarray}
D_1 = \frac{1}{\sqrt2}\left(\begin{array}{c}
-h^+_2+h^+_3 \\
\frac{1}{\sqrt2}(-h^0_2-ia^0_2 +h^0_3+ia^0_3)\end{array}\right),\;\;
D_2 =  \frac{1}{\sqrt2}\left(\begin{array}{c}
h^+_2+h^+_3 \\
\frac{1}{\sqrt2}(h^0_2+ia^0_2 +h^0_3+ia^0_3 )\end{array}\right).
\label{casob}
\end{eqnarray}


\end{document}